\documentclass[newstyle,twocolumn,proceedings]{rmaa}

\usepackage{rmaacite}

\def\hii{H{\sc ii}}
\def\mabs{$M_{\rm B}$}
\def\doh{$12 + \log(\rm O/H)$}

\renewcommand{\P}[1]{%
\ifnum#1=1\hbox{OW~168--326E}\fi
\ifnum#1=2\hbox{OW~167--317}\fi
\ifnum#1=3\hbox{OW~163--317}\fi
\ifnum#1=5\hbox{OW~158--323}\fi
\ifnum#1=0\hbox{OW~171--334}\fi}
\makeatletter
\title{Probing the chemical abundances in distant galaxies with 10-m class telescopes}
\author{Thierry Contini\affil{Observatoire Midi-Pyr\'en\'ees, Toulouse, France} }

\fulladdresses{
\item Thierry Contini: Laboratoire d'Astrophysique de l'Observatoire Midi-Pyr\'en\'ees, UMR 5572, 14 avenue E. Belin, F-31400 Toulouse, France ({\tt contini@ast.obs-mip.fr}).}

\shortauthor{T. Contini}
\shorttitle{Chemical abundances in high-redshift galaxies}


\abstract{%
  The determination of chemical abundances in star-forming galaxies and 
the study of their evolution on cosmological timescales are powerful 
tools for understanding galaxy formation and evolution. This contribution 
presents the latest results in this domain. We show that detailed studies 
of chemical abundances in UV-selected, \hii\ and starburst nucleus galaxies, 
together with the development of new chemical evolution models, put 
strong constraints on the evolutionary stage of these objects in terms 
of star formation history. Finally, we summarize our current knowledge 
on the chemical properties of distant galaxies. 
Although the samples are still too small for statistical studies, 
these results give insight into the nature and evolution of distant 
star-forming objects and their link with present-day galaxies. No doubt 
that the next large-scale spectroscopic surveys on 10-m class telescopes 
will shed light on these fundamental issues.
}
\resumen{La determinaci\'on de abundancias qu\' \i micas en
galaxias con formaci\'on estelar, y el estudio de su evoluci\'on sobre
escalas de tiempo cosmol\'ogicas son piezas clave en la comprensi\'on del
proceso global de formaci\'on y evoluci\'on de las galaxias. Esta
contribuci\'on presenta los \'ultimos resultados en este campo. Se muestra
c\'omo el estudio detallado de abundancias qu\' \i micas en galaxias
seleccionadas en el ultravioleta, ya sean galaxias \hii\ o galaxias con
brotes de formaci\'on estelar nuclear, junto con el desarrollo de nuevos
modelos de evoluci\'on qu\' \i mica, pueden imponer fuertes
restricciones sobre el estado evolutivo de estos objetos en t\'erminos de
su tasa de formaci\'on estelar. Finalmente, se presenta un resumen del
estado actual del conocimiento sobre las propiedades qu\' \i micas de
galaxias lejanas. Aunque las muestras son a\'un demasiado peque\~nas para
estudios estad\' \i sticos, estos resultados proporcionan una primera
visi\'on sobre la naturaleza y la evoluci\'on de las galaxias lejanas con
formaci\'on estelar y su relaci\'on con las galaxias actuales. Sin lugar a
dudas, los futuros surveys espectrosc\'opicos a gran escala sobre
telescopios de la clase de 10-m proporcionar\'an resultados fundamentales
sobre estos temas. 
}

\listofauthors{T.~Contini}
\indexauthor{Contini, T.}

\begin{document}

\maketitle

\section{Introduction}
\label{sec:intro}
Tracing the star formation history (SFH) of galaxies is essential 
for understanding galaxy formation and evolution. The chemical 
properties of galaxies are closely related to their SFH, 
and can be considered as fossil records, enabling us 
to track the galaxy formation history down to the present. 

There is a lot of observational evidence suggesting that 
the SFH of galaxies has not been monotonic
with time, but exhibits instead significant fluctuations. 
Galaxies in the Local Group are excellent examples showing a 
variety of SFHs. 
Further evidence for the ``multiple-burst'' scenario was recently 
found in massive Starburst Nucleus Galaxies (SNBGs; 
\pcite{Cozetal99,Lancetal01}). 
Even the small-mass and less evolved \hii\ galaxies seem to be formed 
of age-composite stellar populations indicating successive bursts of 
SF \cite{Raietal00}. 
\scite{KCB01} recently explored numerical models of galaxy evolution in 
which star
formation occurs in two modes: a low-efficiency continuous mode, and 
a high-efficiency mode triggered by interaction with a
satellite. With these assumptions, the SFH of
low-mass galaxies is characterized by intermittent bursts of SF 
separated by quiescent periods lasting several Gyrs, whereas
massive galaxies are perturbed on time scales of several hundred Myrs
and thus have fluctuating but relatively continuous SFHs. 

Examining the chemical evolution and physical nature of star-forming
galaxies over a range of redshifts will shed light on this issue.
Emission lines from \hii\ regions have long been the primary means of
chemical diagnosis in local galaxies, but this method has only
recently been applied to galaxies at cosmological distances following
the advent of infrared spectrographs on 8 to 10-m class telescopes
\cite{Petetal98,KZ99,KK00,Hametal01,Petetal01} .

\begin{figure}[t]
\includegraphics[width=\columnwidth]{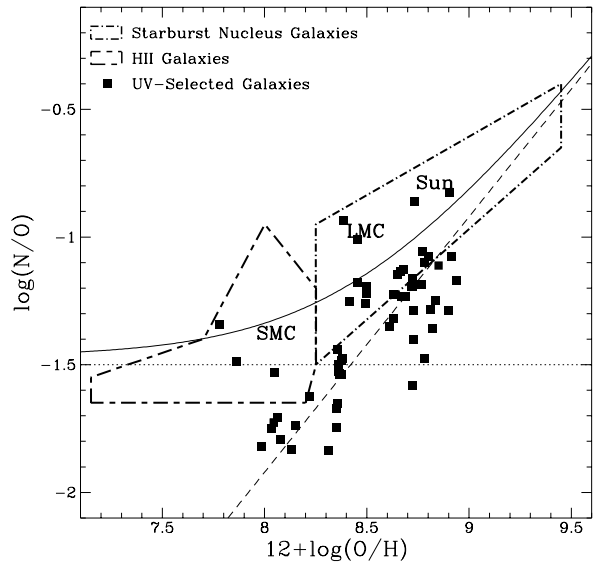}
\caption{N/O versus \doh\ for the
UV-selected galaxies (squares) and two comparison
samples of nearby star-forming galaxies: Starburst Nucleus Galaxies and 
\hii\ galaxies (see \pcite{Contetal02} for references).
}
\label{UV}
\end{figure}

\begin{figure}[t]
\includegraphics[width=\columnwidth]{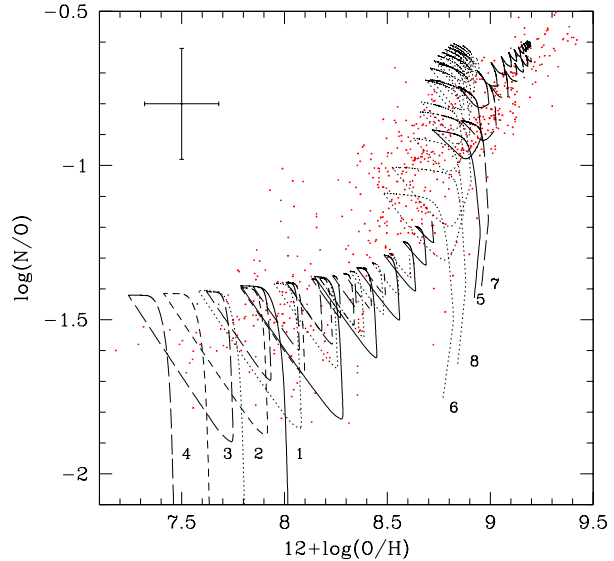}
\caption{N/O versus \doh\ for the samples of star-forming galaxies 
(dots) described in Figure~\ref{UV}. Model predictions assuming 
bursting SF scenario are shown. Model parameters are listed 
in \scite{MC02}. 
}
\label{models}
\end{figure}

\section{The case of UV-selected galaxies} 
\label{sec:constraints}
\scite{Contetal02} recently derived the chemical properties of 
a UV-selected sample of galaxies. These objects are found to be 
intermediate between low-mass, metal-poor \hii\ galaxies and more 
massive, metal-rich SBNGs (see \pcite{Cozetal99} for the dichotomy), 
spanning a wide range of oxygen abundances, from $\sim$ 0.1 to 1 Z$_{\odot}$. 

The behavior of these starburst galaxies in the N/O versus O/H 
relation (see Figure~\ref{UV}) has been investigated in order to probe 
their physical 
nature and SFH (see \pcite{Contetal02} for details). 
At a given metallicity, the majority of UV-selected galaxies 
has low N/O abundance ratios whereas SBNGs show an excess of nitrogen 
abundance when compared to \hii\ regions in the disk of normal
galaxies (see also \pcite{Cozetal99}). The interpretation of these 
behaviors is not straightforward. 
A possible interpretation of the location of UV-selected galaxies and
SBNGs in the N/O versus O/H relation could be that UV galaxies are observed 
at a special stage in their evolution, following a powerful starburst that enriched their ISM in oxygen \cite{Contetal02}, whereas SBNGs 
experienced successive starbursts over the last Gyrs to produce the 
observed nitrogen abundance excess \cite{Cozetal99}.

\section{Constraining the SFH of galaxies}
\label{sec:anal}

\begin{figure}[t]
\includegraphics[width=\columnwidth]{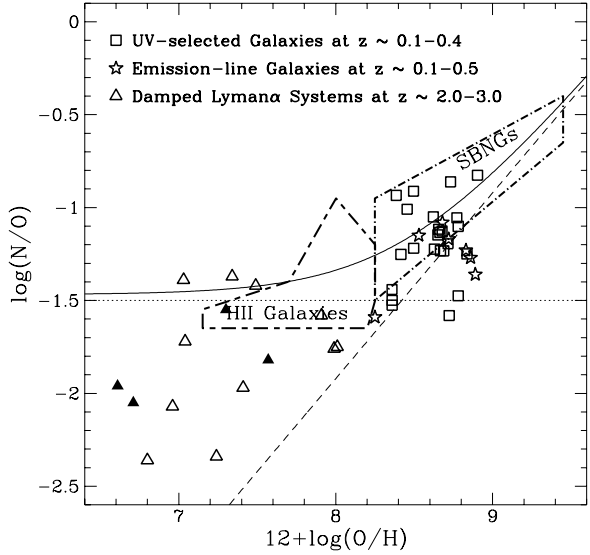}
\caption{N/O as a function of metallicity for distant star-forming 
galaxies. All chemical abundance data published so far for high-z galaxies 
are on this plot (see text for references). 
The location of damped Lyman $\alpha$ systems (triangles; 
\pcite{Petetal02}) at $z \sim 2.0-3.0$ is also shown. 
Filled triangles indicate upper limits. 
}
\label{no_highz}
\end{figure}

\begin{figure}[t]
\includegraphics[width=\columnwidth]{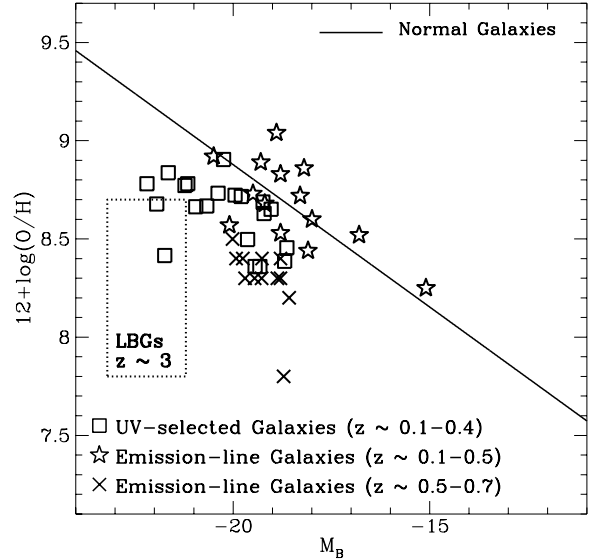}
\caption{Absolute $B$-band magnitude as a function of metallicity for 
distant galaxies (see text for references). 
The location 
of high-redshift ($z \sim 3$) Lyman break galaxies is indicated as a box 
encompassing the range of O/H and \mabs\ derived for 
these objects \cite{Petetal01}. 
}
\label{oh_highz}
\end{figure}

At a given metallicity, the distribution of N/O abundance ratios 
shows a large dispersion, both at low and high metallicity 
\cite{Cozetal99,Contetal02}. 
Only part of this scatter is due to uncertainties in the abundance 
determinations. The additional dispersion must therefore be 
accounted for by galaxy evolution models.
Various hypotheses (localized chemical ``pollution'', IMF variations, 
differential mass loss, etc) were suggested as responsible for such a
scatter (e.g.\,\pcite{KS98}) but none of them is able to 
reproduce the full N/O scatter at a given metallicity. 

A natural explanation for the variation of N/O at constant metallicity 
might be a significant time delay between the release, into the ISM, 
of oxygen by massive, short-lived stars and that of nitrogen produced 
in low-mass longer-lived stars. 
The ``delayed-release'' model assumes that SF is an 
intermittent process in galaxies, while maintaining an universal IMF 
and standard stellar nucleosynthesis \cite{EP78,G90}. 

\scite{MC02} recently investigated this possibility in order to quantifiy 
the SFH of starburst galaxies over a large range of mass and metallicity. 
The observed dispersion in the well-known metallicity-luminosity relation 
has been used as an additional constraint. 
It was confirmed that {\em continuous} SF models are 
unable to reproduce the scatter observed in both N/O and \mabs\ 
versus O/H scaling relations. 
The dispersion associated with the distribution of N/O as a function 
of metallicity can indeed be explained in the framework of {\em bursting} 
SF models. Figure~\ref{models} shows the oscillating behavior 
of the N/O ratio due to the alternating bursting and quiescent phases. 
In this case, the observed dispersion in the N/O versus O/H relation is 
explained by the time delay between the release of oxygen by massive 
stars into the ISM and that of nitrogen by intermediate-mass stars. 
During the starburst events, as massive stars dominate the chemical 
enrichment, the galaxy moves towards the lower right part of the
diagram. During the interburst period, when no SF is 
occurring, the release of N by low and intermediate-mass stars occurs 
a few hundred Myrs after the end of the burst and increases N/O at 
constant O/H. The dilution of interstellar gas by the newly accreted 
intergalactic gas is also observed during the quiescent phases. 

Extensive model computations (see \pcite{MC02} for details) 
show that no possible 
combination of the model parameters (i.e., burst duration, interburst 
period, SF efficiency, gas accretion timescale, etc) is able to account 
for the observed spread for the whole sample of galaxies. 
They found that metal-rich spiral galaxies differ from metal-poor ones by 
a higher SF efficiency and starburst frequency. Low-mas galaxies experienced 
a few bursts of SF whereas massive spiral galaxies experienced numerous 
and extended powerful starbursts.  



\begin{figure}[t]
\includegraphics[width=\columnwidth]{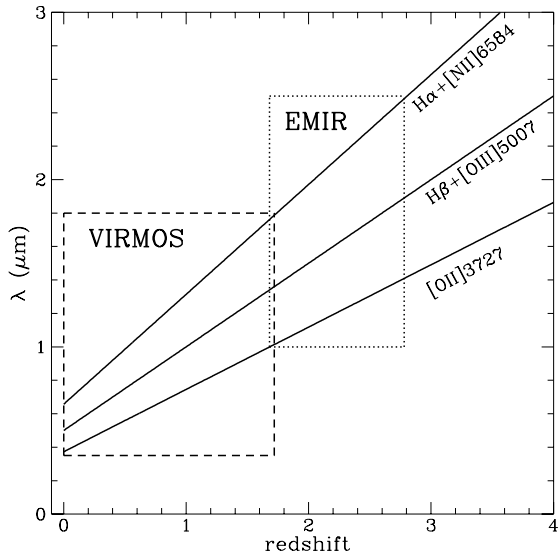}
\caption{Location of the main optical emission lines as a function of 
redshift. With the new generation of visible/infrared multi-object 
spectrographs, such as VIRMOS/VLT and EMIR/GTC, both the O/H and 
N/O abundance ratios will be estimated in thousands of star-forming 
galaxies up to $z \sim 3$.}
\label{elines}
\end{figure}


\section{Physical properties of distant galaxies}
\label{sec:comp}

In Figure~\ref{no_highz}, we compare the location of 
intermediate-redshift ($z \sim 0.1-0.4$) emission-line 
(EL; \pcite{KZ99}) and UV-selected \cite{Contetal02} galaxies with nearby 
samples of \hii\ and SBNGs in the N/O versus O/H plane. 
Most of these distant objects have chemical abundances typical of massive and 
metal-rich SBNGs (i.e., \doh\ $\geq 8.5$). Some of them show high 
N/O abundance ratios (log(N/O) $\ga -1$) which could be due to 
a succession of starbursts over the last few Gyrs (e.g.\,\pcite{Cozetal99}). 
None of these intermediate-redshift galaxies 
are located in the region occupied by nearby metal-poor \hii\ galaxies.
The fact that intermediate-redshift galaxies are mostly metal-rich 
objects is likely to be a selection effect arising 
from the well-known metallicity--luminosity relationship. Only the 
most luminous, and thus metal-rich objects, are detected so far with 
the current instrumentation on 8-m class telescopes. It is interesting to note 
that, like a significant fraction of UV-selected galaxies, some 
optically-selected EL galaxies also show strikingly low N/O ratios.

The location of distant galaxies in the metallicity--luminosity relation 
is shown in Figure~\ref{oh_highz}. Three samples of
intermediate-redshift galaxies are considered: EL galaxies at 
$z \sim 0.1-0.5$ \cite{KZ99}, luminous and 
compact EL galaxies at $z \sim 0.5-0.7$ \cite{Hametal01} and 
the UV-selected galaxies \cite{Contetal02}. 
The location of high-redshift ($z \sim 3$) Lyman break galaxies (LBGs) 
is shown as a box encompassing the range of O/H and \mabs\ derived for 
these objects \cite{Petetal01}. 
Whereas EL galaxies with redshifts between $0.1$ and $0.5$
seem to follow the metallicity--luminosity relation of ``normal'' 
galaxies (solid line), there is a clear deviation for both UV-selected 
and luminous EL galaxies at higher redshift ($z \sim 0.5-0.7$). 
These galaxies thus appear $2-3$ mag brighter than
``normal'' galaxies of similar metallicity, as might be expected
if a strong starburst had temporarily lowered their mass-to-light
ratios. \scite{Hametal01} argue that luminous and compact EL galaxies 
could be the progenitors of present-day spiral
bulges. The deviation is even stronger for LBGs at $z \sim 3$. 
Even allowing for uncertainties in the determination 
of O/H and \mabs, LBGs fall well below the 
metallicity--luminosity relation of ``normal'' local galaxies and 
have much lower abundances than expected from this relation given 
their luminosities. The most obvious interpretation \cite{Petetal01} 
is that LBGs have mass-to-light ratios significantly lower than those 
of present-day ``normal'' galaxies. 

Although the samples are still too small to derive firm conclusions on 
the link between distant objects and present-day galaxies, the present 
results give new insight into the nature and evolution of distant 
star-forming galaxies. No doubt that the next large-scale spectroscopic 
surveys on 10-m class telescopes (e.g.\,VIRMOS on VLT and COSMOS/EMIR on 
Grantecan; see Fig.~\ref{elines}) will shed light on these fundamental 
issues by producing statistically significant samples of galaxies over 
a large range of redshifts.


\end{document}